\documentclass[journal]{IEEEtran}

\usepackage{cite}

\ifCLASSINFOpdf
  \usepackage[pdftex]{graphicx}
  \graphicspath{{fig/}}
  \DeclareGraphicsExtensions{.pdf,.jpeg,.png}
\else
\fi

\usepackage{amsmath,amsfonts,amssymb}
\usepackage{mathtools}
\usepackage{algpseudocode}
\usepackage{algorithm}

\usepackage{array}

\usepackage[utf8x]{inputenc}
\usepackage{bm}
\usepackage{subcaption}	
\usepackage{import}
\usepackage{color}
\usepackage[squaren]{SIunits}
\usepackage{upgreek}
\usepackage{hyperref}

\usepackage{tikz}
\usetikzlibrary{positioning}

\makeatletter
\renewcommand*\env@matrix[1][\arraystretch]{%
  \edef\arraystretch{#1}%
  \hskip -\arraycolsep
  \let\@ifnextchar\new@ifnextchar
  \array{*\c@MaxMatrixCols c}}
\makeatother

\usepackage[opacity=.5]{pdfcomment}

\usepackage[acronym,nohypertypes={acronym}]{glossaries}
\newacronym{ode}{ODE}{ordinary differencial equation}
\newacronym{gp}{GP}{Gaussian process}

\renewcommand{\figurename}{Fig.}
\renewcommand{\tablename}{Tab.}

\newcommand{\capname}{Sec.}

\newcommand{\figref}[1]{\figurename~\ref{#1}}

\newcommand{\capref}[1]{\capname~\ref{#1}}

\newcolumntype{C}[1]{>{\centering\arraybackslash}p{#1}}

\newcommand{\openai}{OpenAI}

\begin{document}
\title{Towards a Scalable and Flexible Simulation and Testing Environment Toolbox for Intelligent Microgrid Control}
\author{\IEEEauthorblockN{Henrik Bode, Stefan Heid, Daniel Weber, Eyke H\"ullermeier, Oliver Wallscheid}\\
\IEEEauthorblockA{Faculty for Computer Science, Electrical Engineering and Mathematics\\
Paderborn University, Germany\\
E-Mail: $\{\,$\href{mailto:henrik.bode@uni-paderborn.de}{henrik.bode}, \href{mailto:stefan.heid@uni-paderborn.de}{stefan.heid}, \href{mailto:daniel.weber@uni-paderborn.de}{daniel.weber}, \href{mailto:eyke.huellermeier@uni-paderborn.de}{eyke.huellermeier}, \href{mailto:oliver.wallscheid@uni-paderborn.de}{oliver.wallscheid}$\,\}$@uni-paderborn.de}
\thanks{The authors would like to thank Jarren Lange from Paderborn University for providing much MSG controller help and Andreas Heuermann from Link\"oping University for support on OpenModelica and FMU integration issues.}}
%

\maketitle
\begin{abstract}
Micro- and smart grids (MSG) play an important role both for integrating renewable energy sources in conventional electricity grids and for providing power supply in remote areas.
Modern MSGs are largely driven by power electronic converters due to their high efficiency and flexibility.
Nevertheless, controlling MSGs is a challenging task due to highest requirements on energy availability, safety and voltage quality within a wide range of different MSG topologies.
This results in a high demand for comprehensive testing of new control concepts during their development phase and comparisons with the state of the art in order to ensure their feasibility.
This applies in particular to data-driven control approaches based on reinforcement learning (RL), the stability and operating behavior of which can hardly be evaluated a priori.
Therefore, the OpenModelica Microgrid Gym (OMG) package, an open-source software toolbox for the simulation and control optimization of MSGs, is proposed.
It is capable of modeling and simulating arbitrary MSG topologies and offers a Python-based interface for plug \& play controller testing.
In particular, the standardized OpenAI Gym interface allows for easy RL-based controller integration.
In addition to presenting the OMG toolbox, several application examples are highlighted, including safe Bayesian optimization for low-level controller tuning.
\end{abstract}


\IEEEpeerreviewmaketitle

\section{Introduction}
\label{sec:intro}
\IEEEPARstart{T}{he} transition of conventional energy supply systems based on fossil fuels to a sustainable structure characterized by renewable energies is a central technical and social challenge of the 21st century \cite{UniNations2020}.
To achieve this, the inherent volatility of renewable energy sources requires a shift away from conventional, centralized structured top-down energy networks towards flexible, cross-sectoral and intelligent energy systems \cite{Lund2017}.
Therefore, in the course of the energy transition, micro- and smart grids (MSG) represent an important solution component to ensure a clean, efficient and cost-effective energy supply\cite{Hatziargyriou2007}\cite{Kroposki2008}.
MSG is the concept of a local network consisting of distributed energy resources (e.g. wind power), energy storage units (e.g. battery) and consumers in various sectors (e.g. electricity, heat, mobility) \cite{Katiraei2006}.
The local integration of renewable energies by means of MSGs, e.g. within industrial companies or residential areas, relieves energy transmission grids and, thus, reduces the need for cost- and resource-intensive grid expansion.
Moreover, MSGs can provide energy supply for remote areas without connection to a public distribution grid.
In this context, power electronic converters became a central component of modern MSG approaches due to their very high energy conversion efficiency and flexibility in order to directly control the power flow between different MSG components \cite{Guerrero2011}. 
 
MSGs are highly heterogeneous, complex systems coming with many different topologies depending on their purpose of application \cite{Hossain2014}\cite{Lidula2011}.
Moreover, their operation contains a significant stochastic component, which is caused by the uncertainty of both the load demand and the regenerative feed-in as well as topology changes due to the insertion or removal of components during operation.
Furthermore, some MSGs may use hybrid AC/DC subgrids in order to boost energy efficiency by reducing the number of required energy conversion stages.
Consequently, controlling MSGs is a demanding task that comes with several key requirements, which can be summarized as follows:
\begin{itemize}
	\item \textbf{Safety}: the continuous availability of energy is of prime importance.
Outages or component failures due to control errors (e.g. by overloading) are unacceptable.
	\item \textbf{Adaptivity}: due to the wide range of MSG use cases, a high degree of control flexibility in a plug \& play sense is necessary.
	\item \textbf{Resource optimality}: minimizing both energy losses and operation costs utilizing available control degrees of freedom is highly desirable. 
	\item \textbf{Power quality}: providing energy supply at high power quality level is important for ensuring nominal functionality at load side. 
\end{itemize}
In order to pursue these objectives, MSG control is typically addressed by hierarchical approaches on different time scales including \cite{Guerrero2013}\cite{Guerrero2013a}:
\begin{itemize}
	\item \textbf{Inner level}: current (and voltage) control in the micro- to millisecond range including auxiliaries such as protective measures or phase-locked loops for each inverter.
	\item \textbf{Primary level}: (re-)active power balancing between different inverters in the (sub-)second range for voltage and/or frequency control.  
	\item \textbf{Secondary level}: energy management (including storage scheduling) focusing on mid-term steady-state correction of important grid key figures (e.g. frequency).
	\item \textbf{Tertiary level}: long-term economic dispatch routines for cost-optimal MSG operation (if one or multiple MSGs contain the necessary operation degrees of freedom). 
\end{itemize}
At the individual levels, the following control approaches can be roughly summarized \cite{Olivares2014}:
\begin{itemize}
	\item Linear feedback controllers such as PID or droop-based characteristics (e.g. \cite{Mohamed2008}\cite{Armin2018}),
	\item (Meta-)heuristic rules and optimization (e.g. \cite{Liu2015}\cite{Khan2017}),
	\item Model predictive control (e.g. \cite{Prodan2014}\cite{Parisio2014}),
	\item Data-driven reinforcement learning (e.g. \cite{Li2012}\cite{Adibi2019}).
\end{itemize}
In most contributions, arbitrary test scenarios are used for the validation of the presented control methods, which are often reduced to a single selected experimental or simulative MSG example.
Obviously, this prevents the comparability of different control procedures on the basis of a common test setup. Moreover, there is a lack of checks whether a method remains functional under different operating conditions or within different MSG topologies.
Hence, there is a high demand for common as well as open test and development platforms to compare MSG control algorithms with each other and to support the development of novel control approaches.

\subsection{Contribution}
\label{subsec:contri}
We present the OpenModelica Microgrid Gym (OMG) package, an open-source software toolbox for the simulation and control optimization of MSGs based on energy conversion by power electronic converters \cite{Bode2020}.
The main contributions and features of the OMG toolbox are:
\begin{itemize}
	\item Flexible and scalable simulations of arbitrary MSG topologies using OpenModelica \cite{OSMC2020} backend;
	\item Python-interface for easy access, configuration and evaluation of arbitrary controllers;
	\item OpenAI gym \cite{Brockman2016} interface for training reinforcement learning agents or similar data-driven approaches;
	\item Single and three phase configurations with AC or DC power supply;
	\item Time domain resolution in the micro- and millisecond range targeting inner and primary level control;
	\item Fully open-source and collaborative project under GNU GPLv3 license.
\end{itemize}

The toolbox is under active development and currently focusing on simulation durations in the second range targeting the inner and primary control level.
Extensions to simplified and lightweight model frameworks for extended simulation horizons are planed. 

Selected background information on implementation of the OMG toolbox are presented in the following and more details can be found in the user guide and API documents \cite{Bode2020}.
Additionally, we present a use case of applying safe Bayesian controller optimization \cite{Berkenkamp2020} to highlight the challenges in the application of data-driven control algorithms with regard to the aforementioned MSG requirements.
In particular, we will deal with the topic of safety in the context of data-driven controller design.
  
\subsection{Related Work}
\label{subsec:related}
In the domain of power system simulations, the following software toolboxes are often mentioned: 
\begin{itemize}
	\item MATPOWER \cite{Zimmerman2011}, an open-source, Matlab-based project targeting static power flow simulation and optimization on distribution grid level.
	Since dynamic modeling is completely omitted, the focus is on secondary and tertiary control level assuming simplified quasi-stationary operation of all components.
	An OpenAI gym interface could be easily added, but is not  available yet. 
	\item Pandapower \cite{Thurner2018}, an open-source, Python-based project targeting static power flow simulation and optimization on distribution grid level.
	It. has a similar scope and functionality as MATPOWER.
	\item PyPSA \cite{Brown2018}, an open-source, Python-based project targeting static power flow simulation and optimization on distributed grid level. It has a similar scope and functionality as the aforementioned packages. 
	\item PSAT \cite{Milano2008}, an open-source, Matlab-based project for simplified single line general power system simulation including optimized scheduling.
	Public user guide or code documentation is not available.
	It comes with a limited, fixed number of pre-defined primary level controllers.
	An external interfacing for other controller types or reinforcement learning is not provided. 
\end{itemize}
Due to the lack both of interfaces and especially dynamic simulation possibilities, the mentioned packages cannot be considered for the control engineering treatment of MSGs on inner and primary control level.
Besides the above mentioned open-source solutions, there is also a range of commercial software with similar functionality focusing on static grid simulations, which is not reported in detail here.
Furthermore, there is a variety of energy market-oriented packages (open-source and commercial) available (e.g. \cite{Balderrana2020}), but since this work is focusing on technical control-oriented problems, these are not discussed here.
In the field of dynamic grid and power electronic simulations, the following software packages have to be mentioned: 
\begin{itemize}
	\item Simscape \cite{Mathworks2020a} is a commercial Matlab/Simulink extension offered by Mathworks.
	It enables a wide range of physics-oriented, dynamic modeling applications including power systems and in particular power electronics.
	Its functionality and scope is similar to the OMG toolbox, but closed-source and interfacing to non-Matlab software products comes with significant calculation overhead (c.f.\ Matlab engine API for Python \cite{Mathworks2020}). 
	\item SPICE-related software such as LTspice, PLECS or ngspice focus on integrated circuit simulations often with nanosecond range time steps.
	Therefore, it is suitable to accurately simulate single power electronic converters on small simulation durations, but computationally not feasible for MSGs with multiple power units.  
\end{itemize}
Therefore, the OMG toolbox is currently the only available open-source solution for dynamic MSG simulations on small time scales.
Due to the offered interfaces, it is particularly suitable for control development and testing, including training and evaluation of recent reinforcement learning techniques.  
\section{Software Description}
\label{sec:software}

\subsection{Toolbox Structure}
\label{subsec:structure}

% Overview and inspiration
The overall structure of the software package is inspired by the tensorforce library\cite{tensorforce}.
OMG contains wrappers for \openai{} Gym environments as well as fully implemented controlling agents.
One of the main contributions of this toolbox is an \openai{} Gym instance in which a reinforcement learning agent can be trained and tested.
However, in order to ease the use of this library, we also provide some predefined agents that can control the environment and also a service class that handles the execution of the specified agent on its environment.
This reduces the boilerplate code significantly and allows developers and engineers to focus on designing and testing controllers.

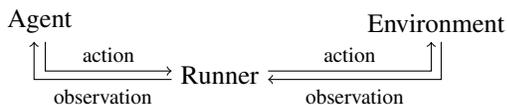
\begin{figure}[h]\centering
	\begin{tikzpicture}[node distance=(.2cm and 1.2cm)]
		\node (runner) {Runner};
		\node[above left=of runner] (agent) {Agent};
		\node[above right=of runner] (env) {Environment};
		
		\draw[->] (runner.5) -| node[above, near start] {\footnotesize action}  (env.-105);
		\draw[->] (env.-75) |- node[below, near end] {\footnotesize observation} (runner.-5);
		\draw[->] (runner.185) -|  node[below, near start] {\footnotesize observation} (agent.-105);
		\draw[->] (agent.-75) |- node[above, near end] {\footnotesize action} (runner.175);
	\end{tikzpicture}
\caption{High level code architecture}\label{fig:arch}
\end{figure}

% features of the runner class
The \textit{runner} class will take care of initializing and termination of agents and environments, as well as the execution of multiple episodes.
The class will handle all information exchange between the agent and the environment, as shown in \figref{fig:arch}.
The functionality is particularly handy, as the training of an agent usually spans multiple training epochs.

% features of the agents
% - wrapper for modifyable controlers
% - 
The \textit{agent} class encapsulates all state related to the learning process.
For example, it may contain a base controller such as a  linear feedback controller that will be parametrized  by external agents during learning and provide the control actions that the agent plays out on the environment (cf. \capref{section:use_case}).
This example corresponds to a hybrid approach mixing expert-driven and data-driven control.
Nevertheless, the definition of the OMG interfaces is completely open, allowing the toolbox to be connected to a wide range of solutions, between entirely data-driven and entirely expert-driven.
The agent also provides possibilities to record monitoring data of the learning process.

% features of the env
% - recorder, plotting
The highly configurable \textit{environment} class provides an interface from OpenAI Gym to the internal simulation model.
It will record data for monitoring and visualizing each epoch, as well as analyzing the control performance in more depth.

\subsection{Modelica Integration}
\label{subsec:Modelica}

\begin{figure*}[h!]
  \begin{center}
 \includegraphics[width=1\linewidth]{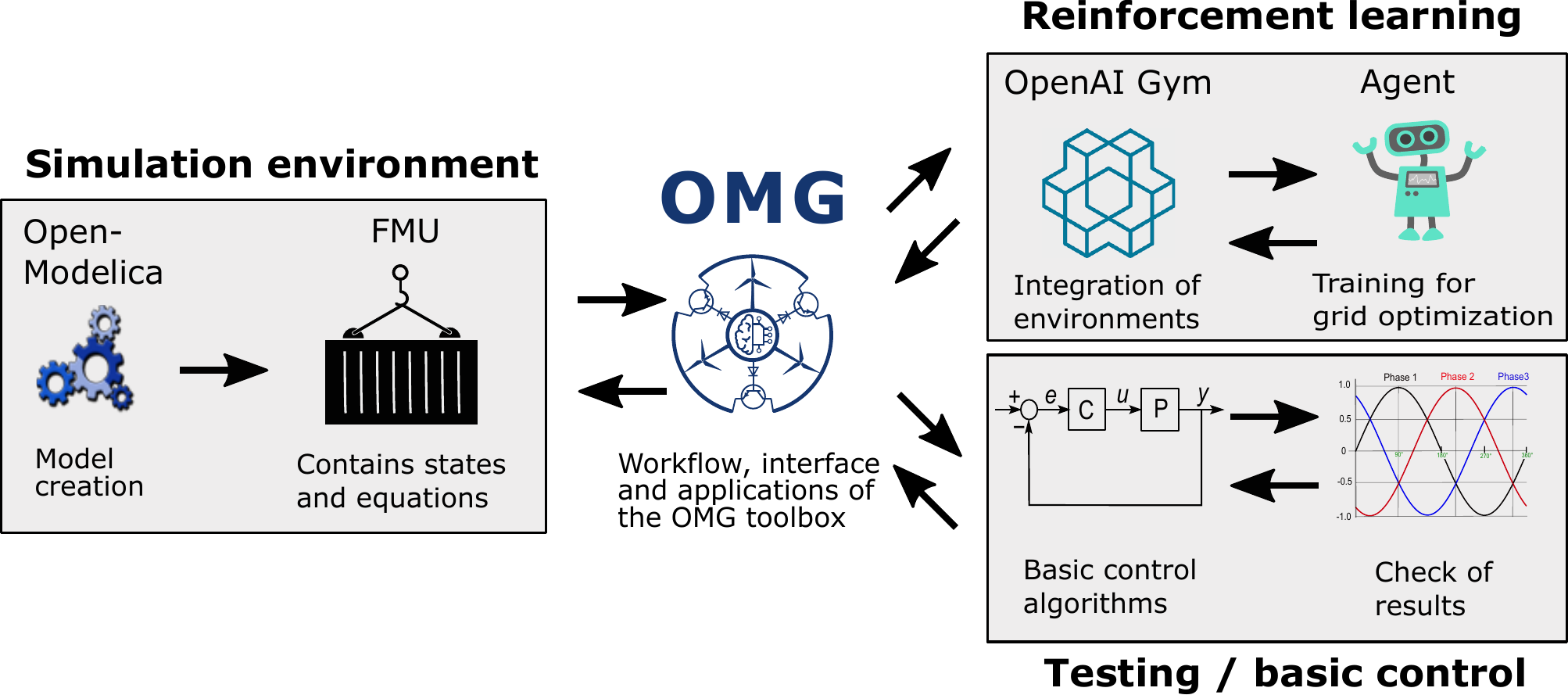}
  \caption{Overview of the interconnections between the different parts of the OMG toolbox. The OpenModelica and OpenAI Gym logos are the property of their respective owners.}
  \label{fig:flowchart}
  \end{center}
\end{figure*}

The overall integration of the OMG toolbox is shown in \figref{fig:flowchart}.
For the model transfer from OpenModelica to Python, the functional mock-up interface (FMI) \cite{fmi} is used.
FMI is a tool-independent open source standard for the exchange of dynamic models.
According to this standard, the model-including objects created for the exchange are called functional mock-up units (FMU), which support two flavours of simulation types: co-simulation (CS) and model exchange (ME).
In CS, the numerical solver is embedded and supplied by the exporting tool, in this case OpenModelica. On the other side, in ME, the importing tool supplies the solver, while the FMU only provides the differential equation system.

Since the \textit{explicit Euler forward} is the only solver implemented for CS in OpenModelica so far, ME is the preferred choice and the default one in OMG.
To avoid small simulation step sizes in order to ensure numerical stability, implicit solvers are required.
Here, the Python package SciPy \cite{solve_ivp} provides several implicit solvers and is used as the standard solution within the OMG toolbox.

The model is imported via PyFMI \cite{pyfmi}, which provides a wide range of communication methods between the model in the FMU and the Python interface.
After extracting the initial states and the equation system, the controllers define the actions (i.e. control input) for the following step.
Next, the equation system gets solved, and the states are transferred to the system interface.

This step-by-step approach increases the simulation time due to an overhead, which is not required by other simulation types, but it provides full access to any result and parameter at any time of the simulation, which is essential for some advanced techniques such as reinforcement learning (RL).

\subsection{Microgrid Modelica Library}
\label{subsec:MG_library}

\begin{figure}[h]
  \begin{center}
 \includegraphics[width=\linewidth]{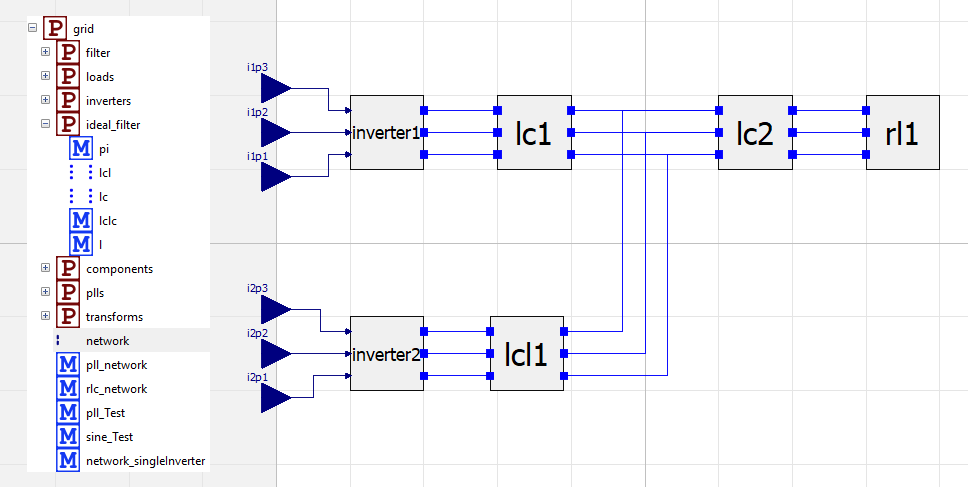}
  \caption{OpenModelica library and an example grid}
  \label{fig:openmodelica}
  \end{center}
\end{figure}

Together with the OMG Python package, an OpenModelica library to create customized MSG topologies is provided. It mainly consists of freely linkable inverters, filters and loads. The library is sketched in \figref{fig:openmodelica} together with an example network.

In this example, a DC bus, which can be adjusted via Python, supplies each inverter. The inverters can be connected via filters (e.g. LC or LCL).
Moreover, a wide range of different load nodes are pre-defined in the library, which can be also be extended by the user.
The filters and loads can be freely parametrized, either directly in the OpenModelica model or via PyFMI. In addition, the toolbox provides auxiliary components (such as phase-locked loops) and pre-defined voltage and current forming inverters, the latter with direct and indirect droop controllers.

Complex loss models for each filter are included.
Due to their large impact on the complexity of the system and the resulting increase of simulation time, it is recommended to use the loss models only if the efficiency and loss behavior is of particular interest.
Switching losses in the inverters are not implemented yet, but can be easily integrated on the Python level of the OMG toolbox.

The controller behavior at load steps is a very important point when addressing MSGs. Due to the lack of structure variability of equation systems in the Modelica language, it is currently not possible to fully add and delete load nodes during a simulation (step-like system topology changes). With further progress in the active research area of multimode modeling, this kind of switching behavior will be implemented in the future. 
In its current version, OpenModelica provides \textit{switches} only in the form of variable resistors having high resistance for an open switch.
Therefore, even open switches are conductive and the high resistance increases the stiffness of the \gls{ode} system, resulting in numerical difficulties of the simulation. 

As a workaround, load steps are provided through parameter-variation of the load nodes.
These parameters can be changed freely at any time during the simulation and, therefore, imitate step-like system changes although in a not entirely physical correct way.

\section{Use Case: Safe Bayesian Optimization for Low-Level Inverter Controller Tuning}
\label{section:use_case}

In the following, the previously introduced OMG toolbox is applied to an exemplary automatic low-level inverter controller tuning process by safe Bayesian optimization to highlight its usage in MSG control scenarios. To this end, we will briefly summarize the fundamentals of inner level inverter current and voltage control in the MSG context, before linking this with a data-driven controller optimization.

\subsection{Rotation Reference Frame - A Short introduction}
MSGs can be described by an \gls{ode} system with state variables $\bm{x}$ like currents and voltages, and control inputs $\bm{u}$ such as the inverter duty cycle:
\begin{equation}
 \dot{\bm{x}} = \bm{f}(\bm{x},\bm{u}) .
\end{equation}
Applying for example an ohmic-inductive load $(R_\mathrm{a}, L_\mathrm{a})$ with a sinusodial voltage $v_\mathrm{a}$ in a (simplified) one-phase grid, the differential equation for controlling the current $i_\mathrm{a}$ is
\begin{equation}
 v_\mathrm{a} = R_\mathrm{a} i_\mathrm{a} + L_\mathrm{a} \frac{\mathrm{d}i_\mathrm{a}}{\mathrm{d}t} \,.
\end{equation}

Extending this for a three-phase system, the state-space model can be described as follows:
\begin{gather}
 \dot{\bm{x}} = \bm{A} \bm{x} + \bm{B} \bm{u}\\
\shortintertext{with}
\begin{aligned}
    \dot{\bm{x}} &=  \begin{bmatrix}
    \dot{i_\mathrm{a}} & \dot{i_\mathrm{b}}& \dot{i_\mathrm{c}}
    \end{bmatrix}^\mathsf{T},&
       \bm{u} & = \begin{bmatrix}
       v_\mathrm{a}& v_\mathrm{b}& v_\mathrm{c}
       \end{bmatrix}^\mathsf{T},
    \\
    \bm{A} & = \begin{bmatrix}
    -\frac{R_\mathrm{a}}{L_\mathrm{a}} & 0 & 0\\
    0 & -\frac{R_\mathrm{b}}{L_\mathrm{b}} & 0\\
    0 & 0 & -\frac{R_\mathrm{c}}{L_\mathrm{c}}
    \end{bmatrix}, &
    \bm{B}  & =  \begin{bmatrix}
    \frac{1}{L_\mathrm{a}}& 0 & 0\\
    0 & \frac{1}{L_\mathrm{b}} & 0\\
    0 & 0 & \frac{1}{L_\mathrm{c}}
\end{bmatrix}.
\end{aligned}\notag
\end{gather}

The current can be represented as a vector in a fixed $ \mathrm{abc} $ reference frame.
This vector is rotating with the frequency of the sinusoidal supply voltage. 
%Due to .... PI controller bad performance in rotation frame.
Using the Park transformation, the system variables can be mapped into a rotating reference frame. 
Here, the $ \mathrm{d} $-axis is aligned with the $ \mathrm{a} $-axis of the rotating three-phase system, the $ \mathrm{q} $-axis is orthogonal to the $ \mathrm{d} $-axis and the third is the zero component:
\begin{equation*}
  \begin{bmatrix} 
    x_\mathrm{d}\\
    x_\mathrm{q}\\
    x_\mathrm{0}
 \end{bmatrix}
 =
 \frac{2}{3}
   \begin{bmatrix}[1.5]
    \mathrm{cos}(\theta) & \mathrm{cos}(\theta - \frac{2\pi}{3}) & \mathrm{cos}(\theta - \frac{4\pi}{3})\\
    -\mathrm{sin}(\theta) & -\mathrm{sin}(\theta - \frac{2\pi}{3}) & -\mathrm{sin}(\theta - \frac{4\pi}{3})\\
    \frac{1}{2} & \frac{1}{2} &\frac{1}{2}
 \end{bmatrix}
  \begin{bmatrix} 
    x_\mathrm{a}\\
    x_\mathrm{b}\\
    x_\mathrm{c}
 \end{bmatrix}.
\end{equation*}

If the angular speed of the rotating frame is set equal to the grid frequency, the sinusoidal grid voltages and currents become stationary DC-variables. 
This simplifies the control design a lot, especially if linear feedback control such as PI controllers is applied.
For more information on the basics of power electronic control we refer to \cite{Mattavelli2006} and similar textbooks. 

For the remaining part of the paper, the simulation of the physical grid system is executed in the $ \mathrm{abc} $ frame within OpenModelica, while the $ \mathrm{dq} $ frame is utilized within the Python-based control parts.

\subsection{Problem Statement}
\label{subsec:problem}

\begin{figure}[h]
  \begin{center}
 \includegraphics[width=\linewidth]{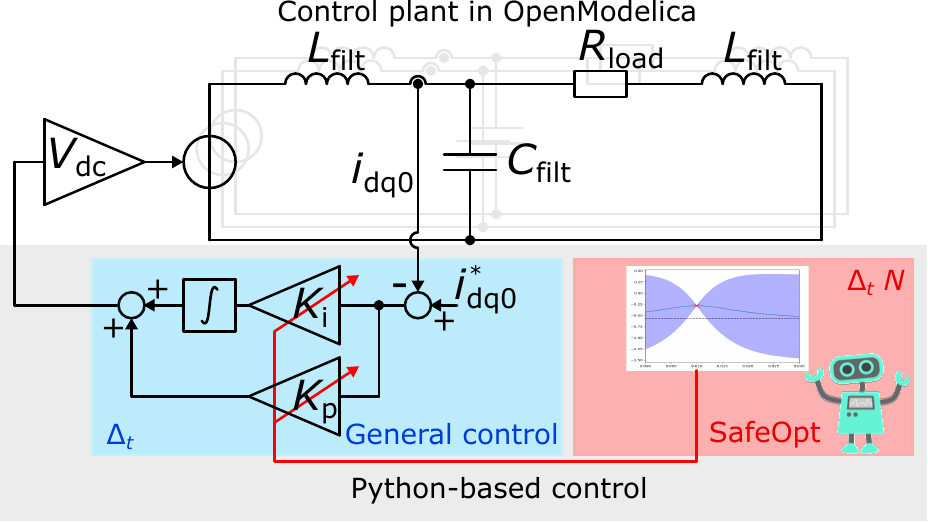}
  \caption{Control plant in OpenModelica in $ \mathrm{abc} $ reference frame and the Python-based control part in $ \mathrm{dq} $ reference frame}
  \label{fig:SI_modell}
  \end{center}
\end{figure}

In this use case an inverter with an LC filter is supplying an ohmic-inductive load. 
As shown in \figref{fig:SI_modell}, standard PI controllers are used to control the currents through the filter inductors in all phases. 
It is assumed that the filter and load parameters are not exactly known and, consequently, the optimal PI controller gains $\{K_i, K_p\}$  cannot be calculated beforehand with respect to a given control performance metric. 
This assumption can be motivated both by parameter variations due to manufacturing uncertainties  \cite{Keyhani2009} and by missing system knowledge from the control point of view, e.g., when the MSG system topology is changed during operation (plug \& play component insertion or removal) \cite{Sadabadi2017}\cite{Riverso2015}.
Consequently, the controller parameters need to be automatically tuned online during regular system operation.
As an important constraint, this tuning has to be performed in a safe way, such that unsuitable gain parameters leading to severe overshoots or other unsafe system behavior are prohibited at all time.

The exemplary use case is to supply the load node by a $15 \, \ampere$ peak current in all three-phases at a grid frequency of $50 \, \hertz$ with zero phase margin. 
The inverter is fed by an idealized DC source of $1000\,\volt$ .
The control is done in the $ \mathrm{dq} $-frame, hence, the setpoints for the current controllers are 
\begin{equation*}
\bm{i}_\mathrm{dq0}^* = [15\, \ampere, 0, 0]^\mathsf{T}	\, ,
\end{equation*}
while for all subsequent training episodes a blackstart is assumed (i.e. all voltages and currents are initially zero). To keep this example simple and clearly arranged, superimposed control loops (such as an additional voltage loop) are not considered, nevertheless, they can be directly integrated into the example files provided in the source code of \cite{Bode2020}. 

For all following investigations, the filter capacity is set as $C_\mathrm{filt} = 20 \, \micro \farad$, the inductance to $L_\mathrm{filt} = 2 \, \milli \henry$, the resistance, the inductance of the load $R_\mathrm{load} = 20 \, \ohm$ and $L_\mathrm{load} = 1 \, \milli \henry$, respectively. These parameters are used for the simulation, but are unknown to the controller tuning processing.

The agent consists of a controller and an optimization part.
In the control part, the observations are used to calculate the modulation indices for the inverter using the PI controllers.
This is done in every step. 
The stepsize is set to $\Delta_t = 50 \, \micro \second$ and the experiment runs for $N= 300$ steps.
As shown in \figref{fig:SI_modell}, the optimization part uses safe Bayesian optimization to calculate new controller parameters after each episode.

\subsection{Solution Approach}
\label{subsec:solution}

The task of the previously described agent is to find optimal controller parameters during online operation. 
To avoid failures during controller optimization, an extension of \textit{Bayesian optimization}\cite{bo-intro}, called \textit{SafeOpt}, has been proposed \cite{Berkenkamp2020}.

In this work, the term \textit{safety} is defined with respect to the performance metric \textit{J} induced by the environments rewards.
Therefore, large negative rewards are seen as indicators of system failure. 
In the defined reward function \eqref{eq:reward}, the mean-root-error (MRE) between the measured phase currents $\bm{i}_\mathrm{abc}$ from the observations and the setpoints $\bm{i}_\mathrm{abc}^*$ is provided as the regular performance indicator. 
However, the MRE is only an exemplary way to evaluate the control performance.
Compared to the classical mean-squared-error (MSE) metric the MRE is penalizing smaller control errors around zero stronger and, therefore, focuses on a steady-state control error free behavior. 
Nevertheless, arbitrary performance measures can be included in the control agent definition within the OMG toolbox. 
Additionally to the MRE, a barrier function is used as a penalty if the nominal current $i_\mathrm{nom}$ is exceeded to avoid that the current limit $i_\mathrm{limit}$ is reached. 
It is assumed that exceeding $i_\mathrm{limit}$ will either lead to severe component damage (e.g. by thermal overloading of the power electronic semiconductors) or to an emergency shutdown of the system. For this example, the nominal current and its limit are set to
\begin{equation*}
	i_\mathrm{nom} = 20\,\ampere, \quad i_\mathrm{limit} = 30\,\ampere\, .
\end{equation*}
The total reward is then given by
\begin{align}
 r = & -\sum_{\mathrm{abc}}\left(\sqrt{\frac{\left|i_\mathrm{abc}^* - i_\mathrm{abc}\right|} {i_\mathrm{limit}}} \right. \nonumber \\
      &- \left. \frac{\left( \mu \cdot \mathrm{log}\left( 1 - \mathrm{max}(|i_\mathrm{abc}| - i_\mathrm{nom}, 0) \right)  \right)}
     {i_\mathrm{limit} - i_\mathrm{nom}}
      \right)
      \label{eq:reward}
\end{align}
with $\mu$ being the weight of the barrier function. 
In the following, $\mu = 2$ is chosen.  
The performance $J$ is calculated as the average reward per episode over all $N$ time steps:
\begin{equation}
 J = \frac{1}{N}\sum_{n = 1}^N r_n \, .
\end{equation}

% what is the save region
The \textit{SafeOpt} algorithm defines safety in a probabilistic manner.
A parameter region is considered as \textit{safe} if the lower confidence bound of the predicted performance does not violate a given lower performance threshold.
The confidence bounds are calculated with a \gls{gp} regression \cite{gpml} over all known performance points seen so far.
This parameter region is called \textit{safe region}.
Consequently, the \textit{SafeOpt} algorithm needs to be initialized with a safe parameter set from which the safe exploration can be started.

% iterative step
% next point maximizes or expands
The parameters considered for subsequent iterations are split into two sets: \textit{expanders} and \textit{potential maximizers}.
\textit{Expanders} are parameters that lay close to the boundaries of the safe region.
Evaluating them increases the knowledge and, therefore, narrows the confidence bounds.
An increased lower confidence bound expands the safe region, because the intersection with the minimal performance bound is pushed out.
\textit{Potential maximizers} are points whose upper confidence bound exceed the currently highest lower bound of the optimal performance.
All other parameters are ignored for evaluation. 
For the next iteration, the parameter set with the largest confidence bound among the two sets is selected.
This solves the exploration/exploitation trade-off in a similar way as the upper confidence bound algorithm \cite{ucb}.

Summarizing, the safety concept relies on two assumptions.
Firstly, the performance in the learning environment must be representative of the real applications faced by the learned agent later on.
This is important, as each parameter set is only evaluated a predefined number of time steps in its learning environment.
Secondly, the \gls{gp} must be able to sufficiently fit the observed performance, otherwise the confidence bounds are not reliable.

A Mat\'{e}rn kernel with $ \nu = 3/2 $ is used as covariance function for the \gls{gp} regression.
Additionally, boundaries for the range of the parameters are specified. 
The \gls{gp} processes thus defined subsequently samples from $\{K_\mathrm{p}, K_\mathrm{i}\}$, which are considered to be \textit{safe} within the estimated confidence bound.

\begin{figure}[h]
  \begin{center}
  \resizebox{\linewidth}{!}{ \hspace{-1.1em} \input{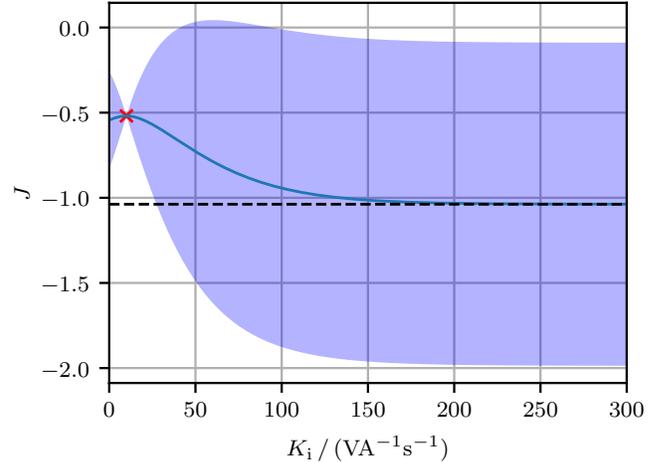}\hspace{-0.2em}}
  \caption{Visualization of the \gls{gp} regression model after one episode. The red marker indicates a parameter evaluation and its performance. The dashed line indicates the chosen safe threshold $J_\mathrm{min}$, the blue curve the mean function and the blue region the $95\,\%$ confidence bounds of the \gls{gp}.}
  \label{fig:GP_oneMeas}
  \end{center}
\end{figure}

\figref{fig:GP_oneMeas} shows a visualization of the \gls{gp} model.
Here, only the control parameter $K_\mathrm{i}$ is considered for the optimization and initialized with $10 \,\volt /(\ampere \second)$, while $ K_\mathrm{p} = 0.01\, \volt / \ampere$ is fixed for demonstration purpose.
The blue curve line shows the mean function of the Gaussian process.
The blue region surrounding the mean function shows the 95\,\% confidence bounds.
Minimal allowed performance, i.e. performances considered as safe, is indicated by the dashed line. 
Therefore, the safe region of $ K_\mathrm{p} $ is  roughly given by $ [0.008\, \volt / \ampere,0.012\, \volt / \ampere]$.

\subsection{Evaluation on Single Gain Tuning of only $K_i$}
\label{subsec:eval_1D}

% 1D Fall: Ki, Beobachtung: Problem lengthscale, 
% Fig 8 schlecht
First, an investigation is carried out assuming a fixed gain $K_p$ while only $K_i$ is tuned by \textit{SafeOpt}. 
We start with this simplified analysis to highlight certain findings, which are much easier to depict when only a single parameter is adapted.
In \capref{subsec:eval_2D}, the evaluation is then extended to a simultaneous tuning of both controller gains.

$K_\mathrm{i}$ is constrained to the range $[0, 300] \, \volt / (\ampere \second)$ and is initialized with $10 \,\volt/(\ampere \second)$. 
This safe initial choice can be motivated by application-specific expert knowledge or robust methods for controller design. 
However, it should be noted that the parameter range limitation is not the safe set.
It potentially contains unsafe regions and can be interpreted as a very vague guess of the optimal location of the controller parameter.

To illustrate the optimization process, $15$ episodes are run by the \textit{SafeOpt} agent.
After each episode, a new parameter for $K_\mathrm{i}$ is chosen.
The algorithm interrupts the episode if the current threshold is exceeded.
The safe threshold $J_\mathrm{min} = 2 \cdot J_\mathrm{init}$ is set to twice the (negative) initial performance $J_\mathrm{init}$.
So, a parameter set is classified as unsafe if the average reward per episode -- indicating the error -- has doubled in comparison to the initial parameter set.

\begin{figure}[h]
  \begin{center}
  \resizebox{\linewidth}{!}{ \hspace{-1.1em} \input{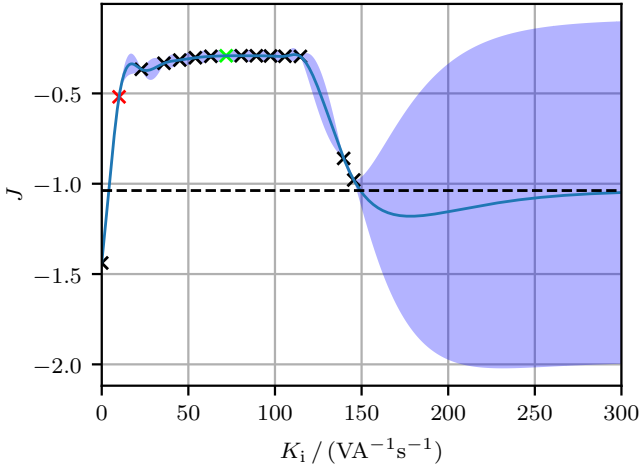}\hspace{-0.3em}}
  \caption{Resulting performance measurement and \gls{gp} model after 15 episodes. Only $K_\mathrm{i}$ is adjusted using \textit{SafeOpt} while $K_\mathrm{p} = 0.005 \,\volt/\ampere$ is kept constant. The dashed line indicates the chosen safe threshold $J_\mathrm{min}$, the blue curve the mean function and the blue region the $95\,\%$ confidence bounds of the \gls{gp}. The red marker represents the initial measurement and the green marker the best.}
  \label{fig:ki_J}
  \end{center}
\end{figure}

In \figref{fig:ki_J}, the results of the \textit{SafeOpt} tuning process is shown, where $K_\mathrm{i}$ is on the x-axis and the performance $J$ on the y-axis. 
The initial performance $J_\mathrm{init} =  -0.52$ is achieved using the above mentioned initial parameter set ($K_\mathrm{p} = 0.005 \, \volt / \ampere$ and $K_\mathrm{i} = 10 \, \volt / (\ampere \second)$).
The plot shows all results of the $15$ performance measurements. 
It can be seen that for increasing integral gain the performance rises until  $K_\mathrm{i} \approx 70\, \volt / (\ampere \second)$ and stays fairly constant until $K_\mathrm{i} \approx 120\, \volt / (\ampere \second)$.
The best performance $J = -0.29$ of the $15$ episodes was achieved using a parameter set of  
\begin{equation*}
 K_\mathrm{p} = 0.005 \, \volt / \ampere \text{ and } K_\mathrm{i} = 71.84 \, \volt / (\ampere \second)\, .
\end{equation*}
If $K_\mathrm{i}$ is increased further, a severe current overshoot exceeding the nominal current limit results during the blackstart of the system. Consequently, the performance decreases significantly. 
If $K_\mathrm{i}$ is decreased relative to the initial value, the stationary control error increases and the performance decreases significantly, too.
The decreasing characteristic for low $K_\mathrm{i}$ already starts from $K_\mathrm{i} \approx 20 \, \volt / (\ampere \second)$. 
This consequently shows that the initial $ K_\mathrm{i} $ was suboptimal.
%The bad performance in \figref{fig:ki_J} is not induced by exceeding the current threshold. \pdfcomment{Hier fehlen Verbindungssätze zum nächsten Abschnitt. Dem Leser wird überhaupt nicht klar, was da jetzt besprochen wird. Selbst ich musste drei mal Lesen bevor ich verstanden haben, dass da jetzt der Fall links aus Fig. 6 besprochen wird. Auch sonst ist der Erklärgehalt in diesem Abschnitt dürftig, bitte mehr Fleisch an den Knochen!}
%

An example for a large stationary control error is shown in \figref{fig:i_abc_ki_J_bad}.
Plotted are the current waveforms achieved by a controller parameter set of $K_\mathrm{p} = 0.005 \, \volt / \ampere$ and $K_\mathrm{i} = 0 \, \volt / (\ampere \second)$.
Since the absolute value of $K_\mathrm{p}$ is low, the reference current cannot be reached. 
Furthermore, since the integral controller gain is zero, the stationary control error is not compensated.
As a result, the performance is unacceptable and the parameter set is classified as unsafe (cf. \figref{fig:ki_J}).
This finding will be discussed in more detail in \capref{subsec:disc}.

\begin{figure}[h]
  \begin{center}
  \resizebox{\linewidth}{!}{\hspace{-1.1em} \input{pictures/inductor_current_bad.pgf} \hspace{-1.1em} }
  \caption{Current waveforms in $\mathrm{abc}$ frame using gain values of $K_\mathrm{p} = 0.005 \, \volt / \ampere$ and $K_\mathrm{i} = 0 \, \volt / (\ampere \second)$}
  \label{fig:i_abc_ki_J_bad}
  \end{center}
\end{figure}

\subsection{Evaluation on Parallel Gain Tuning of $K_p$ and $K_i$}
\label{subsec:eval_2D}

% - zurückgreifen auf J 1D
% - Anzahl der runs anders warum
% - Ref 1D - graue Linie 

\begin{figure}[h!]
  \begin{center}
  \resizebox{\linewidth}{!}{\hspace{-1.1em} \input{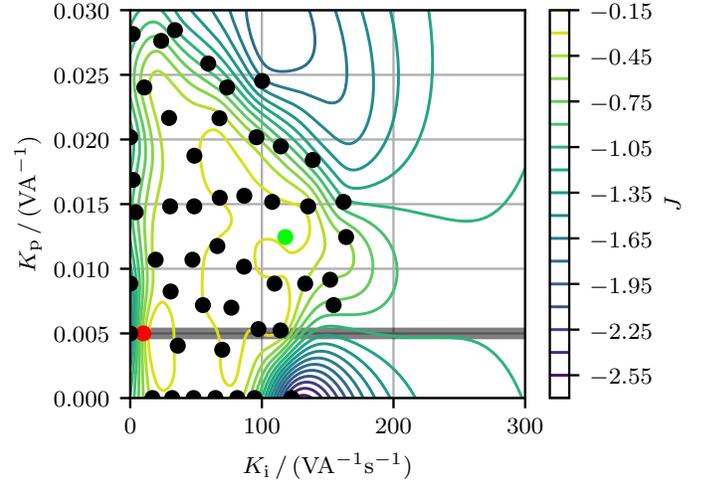}\hspace{-0.8em}}
  \caption{Resulting performance landscape (z-axis) with variable $K_\mathrm{p}$  and $K_\mathrm{i}$. The red marker represents the initial measurement and the green marker the best. The confidence bounds and the safe threshold (set to $J_\mathrm{min} = -1.06$) are not shown. The horizontal grey line at $K_\mathrm{p} = 0.005\, \volt / \ampere$ shows the optimization space of the one dimensional case from \capref{subsec:eval_1D}.}
  \label{fig:kp_ki_J}
  \end{center}
\end{figure}

In the following, both controller parameters are adjusted at the same time to find an optimal controller design.
$K_\mathrm{p}$ is constrained to the range $[0, 0.03]\, \volt / \ampere$ and
$K_\mathrm{i}$ to $[0, 300] \, \volt / (\ampere \second)$.
The initial values are chosen as
\begin{equation*}
		K_\mathrm{p} = 0.005 \,\volt/\ampere \text{ and } K_\mathrm{i} = 10 \,\volt/(\ampere \second) .
\end{equation*}
Using these initial values, the controller reaches a performance of $J_\mathrm{init} = -0.52$. 
As the safe threshold $J_\mathrm{min} = 2 \cdot J_\mathrm{init}$ is selected. 
The \textit{SafeOpt} algorithm is used and the one-dimensional example from \capref{subsec:eval_1D} is extended to two dimensions. 
In \figref{fig:kp_ki_J}, the contour lines visualize the performance landscape.
The integral gain $K_\mathrm{i}$ and proportional gain $K_\mathrm{p}$ are shown on the x- and y-axis, respectively.
The safe threshold $J_\mathrm{min}$ and the confidence bounds are not indicated in this plot.
Since the optimization problem is significantly harder the number of evaluations is increased to $50$.

With respect to the performance landscape in \figref{fig:kp_ki_J}, the parameters found optimal are
\begin{equation*}
		K_\mathrm{p} = 0.0125 \, \volt / \ampere \text{ and } K_\mathrm{i} = 117.81\, \volt / (\ampere \second)\,.
\end{equation*}
This configuration achieves $J =  -0.28$, effectively halving the error compared to the initial performance $J_\mathrm{init}$.
Moreover, this result is slightly better than the performance of the first experiment in \capref{subsec:eval_1D}.
Compared to the optimal parameter values, increasing $K_\mathrm{i}$ further leads to decreasing performance. 
Similarly, increasing $K_\mathrm{p}$ forbids higher values of $K_\mathrm{i}$, because a severe current overshoot exceeding the nominal current limit and stronger oscillations are resulting during the blackstart of the system.
A decreasing $K_\mathrm{p}$ value also results in bad performance for high $K_\mathrm{i}$ values, which causes oscillations initially triggered by the high error at blackstart. 

In \figref{fig:kp_ki_J}, the gray line indicates the constraint parameter space of the first experiments.
The mean function in \figref{fig:ki_J} can be seen as an intersection depicted by the gray line in \figref{fig:kp_ki_J}.
In the second experiment, the gray line is sampled less densely.
Therefore, the model naively predicts quite high performance for $K_\mathrm{i} = 0 \,\volt/(\ampere \second)$.

\figref{fig:i_abc_bk_kp13_Ki135} shows the current waveforms applying the optimal controller design.
Here, a fast control response with only a small current overshoot is achieved while the nominal current is not exceeded.

\begin{figure}[h]
  \begin{center}
  \resizebox{\linewidth}{!}{\hspace{-1.0em} \input{pictures/inductor_current_awesome.pgf}\hspace{-0.2em}}
  \caption{Current waveforms in $\mathrm{abc}$ frame using gain values $K_\mathrm{p} = 0.0125  \, \volt / \ampere$ and ${K}_\mathrm{i} = 117.81 \, \volt / (\ampere \second)$}
  \label{fig:i_abc_bk_kp13_Ki135}
  \end{center}
\end{figure}

\subsection{Discussion}
\label{subsec:disc}
% Discuss problems of GP and this data
In \gls{gp} regression, the fitted function is expected to change equally fast over the whole domain.
This assumption is expressed in the covariance function and its lengthscale.
\figref{fig:ki_J}, however, shows, that such a performance function can be fairly constant in some regions and show rapid change in others.
To allow the \gls{gp} to predict confidence bounds in regions of fast change, the lengthscale of the covariance function has to be chosen extremely small.
This means that parameters that are expected to have strong correlation in their performance values lay very close to each other.
This parametrization, however, also results in very conservative exploration in regions with fairly constant performance.
Unfortunately, the \gls{gp} cannot predict the steep performance drop (cf. again \figref{fig:ki_J}) if neither the prior expectation nor the observed data indicate such a steep change.
Therefore, selecting a larger lengthscale results in highly overconfident behavior as the boundaries of the performance plateau are missed.
Overall the prior covariance assumptions of the \gls{gp} are not met by our performance function, which significantly limits its reliability.
Accordingly, the examined \textit{SafeOpt} algorithm cannot solve the given task satisfactorily. 
In the future, other design methods from the field of safe, data-driven optimization should be examined for their suitability in the context of optimal control of MSGs.

\section{Conclusion and Outlook}
\label{sec:conclusion}
With the OMG toolbox, a fully open-source, scalable and flexible platform for simulation and testing of intelligent microgrid control is proposed. 
The toolbox fills a gap in the area of dynamic system and control analysis for inverter-driven microgrids. 
The core feature is a customizable interface between OpenModelica for plug and play-like system modeling and Python for the integration of arbitrary control algorithms. 
OMG already offers some standard controllers as well as auxiliary tools (e.g.\ phase-looked loops) to speed up the overall simulation and control design process for the user.  
In addition, the integrated OpenAI Gym interface offers a wide range of options for training and evaluating data-driven controllers from the field of reinforcement learning.

The importance of safety has already been highlighted by the data-driven optimization case study of a linear feedback controller. 
Although the standard controller framework is heavily based on expert knowledge and is not a complete data-driven RL agent, its data-driven optimization is associated with the risk of creating unsafe system states, potentially leading to system malfunctions or damages.
Safe Bayesian optimization could only prevent this to a limited extent, because its abstract uncertainty evaluation based on Gaussian processes cannot provide a reliable safety prediction. 

The safe, data-based control of microgrids therefore remains an exciting research challenge for future work. 
Here, the integration of a priori expert knowledge for the evaluation of safe control methods appears to be especially promising, for example to monitor and guide the training of reinforcement learning-based methods.

\subsection*{Acknowledgments}
The authors kindly acknowledge the funding and support of this work by the Paderborn University research grant.				 

\bibliographystyle{IEEEtran}
\bibliography{refs}

\begin{thebibliography}{10}
\providecommand{\url}[1]{#1}
\csname url@samestyle\endcsname
\providecommand{\newblock}{\relax}
\providecommand{\bibinfo}[2]{#2}
\providecommand{\BIBentrySTDinterwordspacing}{\spaceskip=0pt\relax}
\providecommand{\BIBentryALTinterwordstretchfactor}{4}
\providecommand{\BIBentryALTinterwordspacing}{\spaceskip=\fontdimen2\font plus
\BIBentryALTinterwordstretchfactor\fontdimen3\font minus
  \fontdimen4\font\relax}
\providecommand{\BIBforeignlanguage}[2]{{%
\expandafter\ifx\csname l@#1\endcsname\relax
\typeout{** WARNING: IEEEtran.bst: No hyphenation pattern has been}%
\typeout{** loaded for the language `#1'. Using the pattern for}%
\typeout{** the default language instead.}%
\else
\language=\csname l@#1\endcsname
\fi
#2}}
\providecommand{\BIBdecl}{\relax}
\BIBdecl

\bibitem{UniNations2020}
\BIBentryALTinterwordspacing
{United Nations}, ``{Sustainable Development Goals},'' 2020. [Online].
  Available: \url{https://sustainabledevelopment.un.org/?menu=1300}
\BIBentrySTDinterwordspacing

\bibitem{Lund2017}
H.~Lund, P.~A. {\O}stergaard, D.~Connolly, and B.~V. Mathiesen, ``{Smart Energy
  and Smart Energy Systems},'' \emph{Energy}, vol. 137, pp. 556 -- 565, 2017.

\bibitem{Hatziargyriou2007}
N.~{Hatziargyriou}, H.~{Asano}, R.~{Iravani}, and C.~{Marnay}, ``Microgrids,''
  \emph{IEEE Power and Energy Magazine}, vol.~5, no.~4, pp. 78--94, 2007.

\bibitem{Kroposki2008}
B.~{Kroposki}, R.~{Lasseter}, T.~{Ise}, S.~{Morozumi}, S.~{Papathanassiou}, and
  N.~{Hatziargyriou}, ``{Making Microgrids Work},'' \emph{IEEE Power and Energy
  Magazine}, vol.~6, no.~3, pp. 40--53, 2008.

\bibitem{Katiraei2006}
F.~{Katiraei} and M.~R. {Iravani}, ``{Power Management Strategies for a
  Microgrid With Multiple Distributed Generation Units},'' \emph{IEEE
  Transactions on Power Systems}, vol.~21, no.~4, pp. 1821--1831, 2006.

\bibitem{Guerrero2011}
J.~M. {Guerrero}, J.~C. {Vasquez}, J.~{Matas}, L.~G. {de Vicuna}, and
  M.~{Castilla}, ``{Hierarchical Control of Droop-Controlled AC and DC
  Microgrids: A General Approach Toward Standardization},'' \emph{IEEE
  Transactions on Industrial Electronics}, vol.~58, no.~1, pp. 158--172, 2011.

\bibitem{Hossain2014}
E.~Hossain, E.~Kabalci, R.~Bayindir, and R.~Perez, ``{Microgrid Testbeds Around
  the World: State of Art},'' \emph{Energy Conversion and Management}, vol.~86,
  pp. 132 -- 153, 2014.

\bibitem{Lidula2011}
N.~Lidula and A.~Rajapakse, ``{Microgrids Research: A Review of Experimental
  Microgrids and Test Systems},'' \emph{Renewable and Sustainable Energy
  Reviews}, vol.~15, no.~1, pp. 186 -- 202, 2011.

\bibitem{Guerrero2013}
J.~M. {Guerrero}, M.~{Chandorkar}, T.~{Lee}, and P.~C. {Loh}, ``{Advanced
  Control Architectures for Intelligent Microgrids, Part I: Decentralized and
  Hierarchical Control},'' \emph{IEEE Transactions on Industrial Electronics},
  vol.~60, no.~4, pp. 1254--1262, 2013.

\bibitem{Guerrero2013a}
J.~M. {Guerrero}, P.~C. {Loh}, T.~{Lee}, and M.~{Chandorkar}, ``{Advanced
  Control Architectures for Intelligent Microgrids, Part II: Power Quality,
  Energy Storage, and AC/DC Microgrids},'' \emph{IEEE Transactions on
  Industrial Electronics}, vol.~60, no.~4, pp. 1263--1270, 2013.

\bibitem{Olivares2014}
D.~E. {Olivares}, A.~{Mehrizi-Sani}, A.~H. {Etemadi}, C.~A. {Canizares},
  R.~{Iravani}, M.~{Kazerani}, A.~H. {Hajimiragha}, O.~{Gomis-Bellmunt},
  M.~{Saeedifard}, R.~{Palma-Behnke}, G.~A. {Jimenez-Estevez}, and N.~D.
  {Hatziargyriou}, ``{Trends in Microgrid Control},'' \emph{IEEE Transactions
  on Smart Grid}, vol.~5, no.~4, pp. 1905--1919, 2014.

\bibitem{Mohamed2008}
Y.~A.~I. {Mohamed} and E.~F. {El-Saadany}, ``{Adaptive Decentralized Droop
  Controller to Preserve Power Sharing Stability of Paralleled Inverters in
  Distributed Generation Microgrids},'' \emph{IEEE Transactions on Power
  Electronics}, vol.~23, no.~6, pp. 2806--2816, 2008.

\bibitem{Armin2018}
M.~Armin, P.~N. Roy, S.~K. Sarkar, and S.~K. Das, ``{LMI-Based Robust PID
  Controller Design for Voltage Control of Islanded Microgrid},'' \emph{Asian
  Journal of Control}, vol.~20, no.~5, pp. 2014--2025, 2018.

\bibitem{Liu2015}
N.~{Liu}, Q.~{Chen}, J.~{Liu}, X.~{Lu}, P.~{Li}, J.~{Lei}, and J.~{Zhang}, ``{A
  Heuristic Operation Strategy for Commercial Building Microgrids Containing
  EVs and PV System},'' \emph{IEEE Transactions on Industrial Electronics},
  vol.~62, no.~4, pp. 2560--2570, 2015.

\bibitem{Khan2017}
B.~{Khan} and P.~{Singh}, ``{Selecting a Meta-Heuristic Technique for Smart
  Micro-Grid Optimization Problem: A Comprehensive Analysis},'' \emph{IEEE
  Access}, vol.~5, pp. 13\,951--13\,977, 2017.

\bibitem{Prodan2014}
I.~Prodan and E.~Zio, ``{A Model Predictive Control Framework for Reliable
  Microgrid Energy Management},'' \emph{International Journal of Electrical
  Power \& Energy Systems}, vol.~61, pp. 399 -- 409, 2014.

\bibitem{Parisio2014}
A.~{Parisio}, E.~{Rikos}, and L.~{Glielmo}, ``{A Model Predictive Control
  Approach to Microgrid Operation Optimization},'' \emph{IEEE Transactions on
  Control Systems Technology}, vol.~22, no.~5, pp. 1813--1827, 2014.

\bibitem{Li2012}
F.-D. Li, M.~Wu, Y.~He, and X.~Chen, ``{Optimal Control in Microgrid Using
  Multi-Agent Reinforcement Learning},'' \emph{ISA Transactions}, vol.~51,
  no.~6, pp. 743 -- 751, 2012.

\bibitem{Adibi2019}
M.~Adibi and J.~van~der Woude, ``{A Reinforcement Learning Approach for
  Frequency Control of Inverted-Based Microgrids},'' \emph{IFAC-PapersOnLine},
  vol.~52, no.~4, pp. 111 -- 116, 2019.

\bibitem{Bode2020}
\BIBentryALTinterwordspacing
H.~Bode, S.~Heid, D.~Weber, and O.~Wallscheid, ``{OpenModelica Microgrid Gym
  (OMG)},'' 2020. [Online]. Available:
  \url{https://github.com/upb-lea/openmodelica-microgrid-gym}
\BIBentrySTDinterwordspacing

\bibitem{OSMC2020}
\BIBentryALTinterwordspacing
{Open Source Modelica Consortium (OSMC)}, ``{OpenModelica},'' 2020. [Online].
  Available: \url{https://www.openmodelica.org/}
\BIBentrySTDinterwordspacing

\bibitem{Brockman2016}
\BIBentryALTinterwordspacing
G.~Brockman, V.~Cheung, L.~Pettersson, J.~Schneider, J.~Schulman, J.~Tang, and
  W.~Zaremba, ``{OpenAI Gym},'' 2016. [Online]. Available:
  \url{http://arxiv.org/abs/1606.01540}
\BIBentrySTDinterwordspacing

\bibitem{Berkenkamp2020}
\BIBentryALTinterwordspacing
F.~Berkenkamp, ``{SafeOpt: Safe Bayesian Optimization},'' 2020. [Online].
  Available: \url{https://github.com/befelix/SafeOpt}
\BIBentrySTDinterwordspacing

\bibitem{Zimmerman2011}
R.~D. {Zimmerman}, C.~E. {Murillo-Sanchez}, and R.~J. {Thomas}, ``{MATPOWER:
  Steady-State Operations, Planning, and Analysis Tools for Power Systems
  Research and Education},'' \emph{IEEE Transactions on Power Systems},
  vol.~26, no.~1, pp. 12--19, 2011.

\bibitem{Thurner2018}
L.~{Thurner}, A.~{Scheidler}, F.~{Sch\"afer}, J.~{Menke}, J.~{Dollichon},
  F.~{Meier}, S.~{Meinecke}, and M.~{Braun}, ``{Pandapower: An Open-Source
  Python Tool for Convenient Modeling, Analysis, and Optimization of Electric
  Power Systems},'' \emph{IEEE Transactions on Power Systems}, vol.~33, no.~6,
  pp. 6510--6521, 2018.

\bibitem{Brown2018}
T.~Brown, J.~H\"orsch, and D.~Schlachtberger, ``{PyPSA: Python for Power System
  Analysis},'' \emph{Journal of Open Research Software}, vol.~6, no.~4, 2018.

\bibitem{Milano2008}
F.~{Milano}, L.~{Vanfretti}, and J.~C. {Morataya}, ``{An Open Source Power
  System Virtual Laboratory: The PSAT Case and Experience},'' \emph{IEEE
  Transactions on Education}, vol.~51, no.~1, pp. 17--23, 2008.

\bibitem{Balderrana2020}
\BIBentryALTinterwordspacing
S.~Balderrana and S.~Quoilin, ``Micro-grids,'' 2020. [Online]. Available:
  \url{https://github.com/squoilin/MicroGrids}
\BIBentrySTDinterwordspacing

\bibitem{Mathworks2020a}
\BIBentryALTinterwordspacing
Mathworks, ``Simscape,'' 2020. [Online]. Available:
  \url{https://www.mathworks.com/products/simscape.html}
\BIBentrySTDinterwordspacing

\bibitem{Mathworks2020}
\BIBentryALTinterwordspacing
------, ``{MATLAB Engine API for Python},'' 2020. [Online]. Available:
  \url{https://de.mathworks.com/help/matlab/matlab_external/install-the-matlab-engine-for-python.html}
\BIBentrySTDinterwordspacing

\bibitem{tensorforce}
\BIBentryALTinterwordspacing
A.~Kuhnle, M.~Schaarschmidt, and K.~Fricke, ``Tensorforce: a tensorflow library
  for applied reinforcement learning,'' 2017. [Online]. Available:
  \url{https://github.com/tensorforce/tensorforce}
\BIBentrySTDinterwordspacing

\bibitem{fmi}
\BIBentryALTinterwordspacing
{Modelica Association}, ``Functional mock-up interface standard.'' [Online].
  Available: \url{https://fmi-standard.org/}
\BIBentrySTDinterwordspacing

\bibitem{solve_ivp}
\BIBentryALTinterwordspacing
{The SciPy community}, ``scipy.integrate.solve\_ivp.'' [Online]. Available:
  \url{https://docs.scipy.org/doc/scipy/reference/generated/scipy.integrate.solve\_ivp.html}
\BIBentrySTDinterwordspacing

\bibitem{pyfmi}
\BIBentryALTinterwordspacing
{Modelon}, ``Pyfmi.'' [Online]. Available:
  \url{https://pypi.org/project/PyFMI/}
\BIBentrySTDinterwordspacing

\bibitem{Mattavelli2006}
S.~{Buso} and P.~{Mattavelli}, \emph{Digital Control in Power Electronics},
  2006.

\bibitem{Keyhani2009}
A.~{Keyhani}, M.~{Marwali}, and M.~{Dai}, \emph{Integration of Green and
  Renewable Energy in Electric Power Systems}, 2009.

\bibitem{Sadabadi2017}
M.~S. {Sadabadi}, Q.~{Shafiee}, and A.~{Karimi}, ``{Plug-and-Play Voltage
  Stabilization in Inverter-Interfaced Microgrids via a Robust Control
  Strategy},'' \emph{IEEE Transactions on Control Systems Technology}, vol.~25,
  no.~3, pp. 781--791, 2017.

\bibitem{Riverso2015}
S.~{Riverso}, F.~{Sarzo}, and G.~{Ferrari-Trecate}, ``{Plug-and-Play Voltage
  and Frequency Control of Islanded Microgrids With Meshed Topology},''
  \emph{IEEE Transactions on Smart Grid}, vol.~6, no.~3, pp. 1176--1184, 2015.

\bibitem{bo-intro}
B.~Shahriari, K.~Swersky, Z.~Wang, R.~P. Adams, and N.~de~Freitas, ``Taking the
  human out of the loop: A review of {B}ayesian optimization,''
  \emph{Proceedings of the IEEE}, vol. 104, pp. 148--175, 2016.

\bibitem{gpml}
C.~Rasmussen, \emph{Gaussian Processes for Machine Learning}, 2009.

\bibitem{ucb}
P.~Auer, ``Using confidence bounds for exploitation-exploration trade-offs,''
  \emph{Journal of Machine Learning Research}, vol.~3, no. Nov, pp. 397--422,
  2002.

\end{thebibliography}

\end{document}